\documentclass[final,3p,times,twocolumn]{elsarticle}
\usepackage{graphicx}
\usepackage{subcaption}
\usepackage{siunitx}
\usepackage{amssymb}
\usepackage{amsthm}
\usepackage{amsmath}
\usepackage{lineno}
\usepackage{notes2bib}
\niibsetup{writekey=false,name={}}
\bibnotesetup{
  note-name = ,
  use-sort-key = false
}

\usepackage{booktabs} 
\newcommand{\ra}[1]{\renewcommand{\arraystretch}{#1}}
\biboptions{sort&compress}

\journal{(and accepted as per 10.Feb.2020) by Journal of Molecular Liquids}

\begin{document}

\begin{frontmatter}


\title{Diffusion of single active-dipolar cubes in applied fields}

\author[First]{Martin Kaiser$^\ast$}
\ead{martin.kaiser@univie.ac.at}
\author[Second]{Yeimy Martinez}
\author[Second]{Annette M. Schmidt}
\author[Third]{Pedro A. S\'anchez}
\author[Fourth,First]{Sofia S. Kantorovich}

\address[First]{Faculty of Physics, University of Vienna, Boltzmanngasse 5, 1090 Vienna, Austria}
\address[Second]{Chemistry Department, University of Cologne, D-50939 Cologne, Germany}
\address[Third]{Helmholtz-Zentrum Dresden-Rossendorf, Bautzner Landstrasse 400, 01328 Dresden, Germany}
\address[Fourth]{Ural Federal University, Lenin Av. 51, Ekaterinburg 620000, Russian Federation}

\begin{abstract}

``Active matter'' refers to a class of out-of-equilibrium systems whose ability to transform  environmental energy to kinetic energy is sought after in multiple fields of science and at very different length scales. At microscopic scales, an important challenge lies in overpowering the particles reorientation due to thermal fluctuations, especially in nano-sized systems, to create non-random, directed motion, needed for a wide range of possible applications. In this article, we employ molecular dynamics simulations to show that the diffusion of a self-propelling dipolar nanocube can be enhanced in a pre-defined direction with the help of a moderately strong applied magnetic field, overruling the effect of the thermal fluctuations. Furthermore, we show that the direction of diffusion is given by the orientation of the net internal magnetisation of the cube. This can be used to determine experimentally the latter in synthetically crafted active cobalt ferrite nanocubes.
\end{abstract}


\end{frontmatter}

\section{Introduction}
The term ``active matter'' is a broad concept that accounts for any system of one or many entities that convert environmental energy into kinetic energy, hence self-propelled motion. Importantly, such general definition applies to any length scale. A paradigmatic example is the case of animals, that are active entities whose characteristic sizes span from the macroscopic to the microscopic scale. For the latter case, one can find not only other active biological entities, such as bacteria or that biopolymers forming the cytoskeleton of living cells, but also artificial systems created with nano- and molecular motors on the nanoscale. Available experimental techniques have already allowed to explore multiple mechanisms to create active microscopic systems, particularly artificial ``active particles'' that self-propel across an appropriate or taylored environment. Examples of such environments are as diverse as concentration gradients of chemical reactants \cite{velegol2016origins,bechinger2016active} or the use of defocused lasers to induce self-thermophoresis \cite{piazza2008thermophoresis,yang2011simulations,ghosh2009controlled}.

The interest in designed active micro- and nanoparticles has largely grown in recent years due to their great potential for multiple applications. The relevant property one harness from such active particles is their enhanced kinetic energy compared to simple Brownian particles, leading to fascinating applications such as dynamically changing crystal lattices \cite{cates2009lattice,whitelam2018phase,lavrentovich2016active} or targeted drug delivery systems \cite{bonacucina2009colloidal,barbe2004silica,singh2009nanoparticle,needleman2017active}. However, many of such applications require the active component to move within liquid backgrounds that can not be easily taylored while facing the effects of thermal fluctuations, whose relative importance tends to grow as the characteristic length scale of the system decreases. Rotational diffusion becomes a challenge for the design of certain applications, as it tends to make the direction of the active motion unpredictable, particularly at the nanoscale \cite{yadav2015anatomy,santiago2018nanoscale}. Therefore, to find new ways to control and direct the motion of active micro- and nanoparticles, preferably by means of easily applicable external stimuli, is a key aspect for the development of applications based on these systems. One of the most interesting strategies to achieve such a control is based on the use of external magnetic fields. The fact that most biological materials have a negligible response to not very strong fields avoids undesired side effects, making this approach particularly appealing for biomedical applications. The main challenge is the design of artificial active particles with a magnetic response that optimises the external control of their motion.

The idea of gaining control on active systems by designing magnetoresponsive active particles is not uncharted waters. There have been several attempts of using external fields to orient the motion of active particles self-propelled by catalytic reactions, for instance by partially covering their surface with thin layers of magnetic materials \cite{baraban2012catalytic,baraban2013control}. Rotating and non-uniform magnetic fields can be employed to not only orient but also to power the self-propulsion of magnetic particles \cite{cheang2014minimal,ghosh2009controlled,lalande2010experimental}, that become swimming units. However, despite the promising properties that magnetic particles have regarding their field induced reorientation, to taylor and optimise their behaviour as active entities is still an open challenge. To this respect, any rationale design has to take into account the current knowledge on the properties of magnetic colloids and nanoparticles.

Independently from their potential as active materials, micro- and nanoparticles composed entirely or partially of magnetic substances have been studied in the past 60 years as the main building blocks of magnetic soft matter systems. Classical examples of magnetic soft matter are ferrofluids and magneto-rheological fluids, that are suspensions of such magnetic particles in magneto-passive liquid carriers \cite{hiergeist1999application,huang2011study,odenbach2009colloidal,2011-devicente}, as well as magnetic gels \cite{1995-shiga, 1997-zrinyi, 2015-roeder} and elastomers \cite{1996-jolly, 2016-odenbach}. The behaviour of all these materials strongly depends on the self-assembly and field-induced assembly processes of the magnetic particles forming them \cite{2004-klokkenburg, 2007-klokkenburg, 2007-filipcsei, 2012-borbath}. Such assembly properties are determined by the existence of a remanent or field-induced magnetic moment within the particles, with a persistent or changing orientation with respect to their body frame. In the simplest case, corresponding to monocrystalline spherical particles, the magnetisation orientation mainly depends on internal anisotropies associated to the crystalline structure of the material \cite{buschow2003physics,cullity2011introduction}. For anisometric (\textit{i.e.}, non spherical) particles, their shape anisotropy may cause additionally an imbalance in the demagnetising fields, generating a preferential axis for magnetisation orientation. The importance of this interplay between internal and shape anisotropies is growing with the development of modern synthesis techniques, that allow reproducible manufacturing of micro- and nanoparticles with a broad range of anisotropy combinations and magnetic behaviours \cite{Kovalenko2007,rossi11a,Ahniyaz2007,Disch2011,Meijer2013,yan10a,gunter11a,nakade07a,2013-sacanna,tierno14,2015-donaldson-jmmm}.

Among novel anisometric magnetic particles, magnetic nanocubes are currently attracting a considerable attention due to the strong dependence of their properties on the relative orientation of their magnetic preferential axis and their anisometry \cite{liu2005shape, 2007-xiong, rossi2011cubic, 2012-szyndler, 2014-wetterskog, rossi2018self, 2019-li-nl}. Such diversity of behaviours points magnetic nanocubes as extremely interesting test systems for the fundamental study of the impact of anisotropies on active nanoparticles, opening up the possibility to find novel and/or optimised configurations for practical applications.

In this work we present a preliminary qualitative study based on computer simulations of a recently synthesised hybrid active-magnetic nanoparticles. The system is composed of a catalytic active spherical nanoparticle and a single-domain magnetic nanocube rigidly assembled \cite{inprep}. The propulsion and the magnetic axes in this hybrid particle are not necessarily co-aligned, hence very different behaviours can be obtained depending on their relative orientation. Employing a coarse-grained model that accounts for the shape of each component and represents the magnetisation of the cube as a point dipole fixed in its body frame, we perform extensive molecular dynamics simulations of a single active unit, sampling two distinct orientations of the magnetisation in an experimentally accessible range of applied magnetic fields. By thoroughly comparing the active diffusion properties for both investigated cases, we show that an applied field of moderate strength can effectively direct the active diffusion of these particles. We also show that the intrinsic orientation of the magnetisation is the decisive factor that determines the direction of the active diffusion through a measure accessible in experiments. Our results indicate that this effect could be used to determine experimentally the actual magnetic axis of any individual active nanoparticle of this type.

The structure of the article is the following. In the next Section \ref{sec:amc} we introduce the system under study. In Section \ref{subs:exp} we describe the experimental system, whereas Section \ref{subs:sim} includes details of the simulation approach. In \ref{subs:params}, we discuss experimental and model parameters. Section \ref{ref:rnd} presents the results, first discussing the diffusion coefficients for both studied cases in \ref{subs:msd} and finally the impact of magnetisation orientation on the direction of diffusion in \ref{subs:ang}. Final conclusions are summarised in Section \ref{sec:con}.

\section{Active magnetic nanocube} \label{sec:amc}

\subsection{Experimental system}\label{subs:exp}

\begin{figure}
\centering
\begin{subfigure}{0.5\columnwidth}
  \centering
  \includegraphics[width=1.\columnwidth]{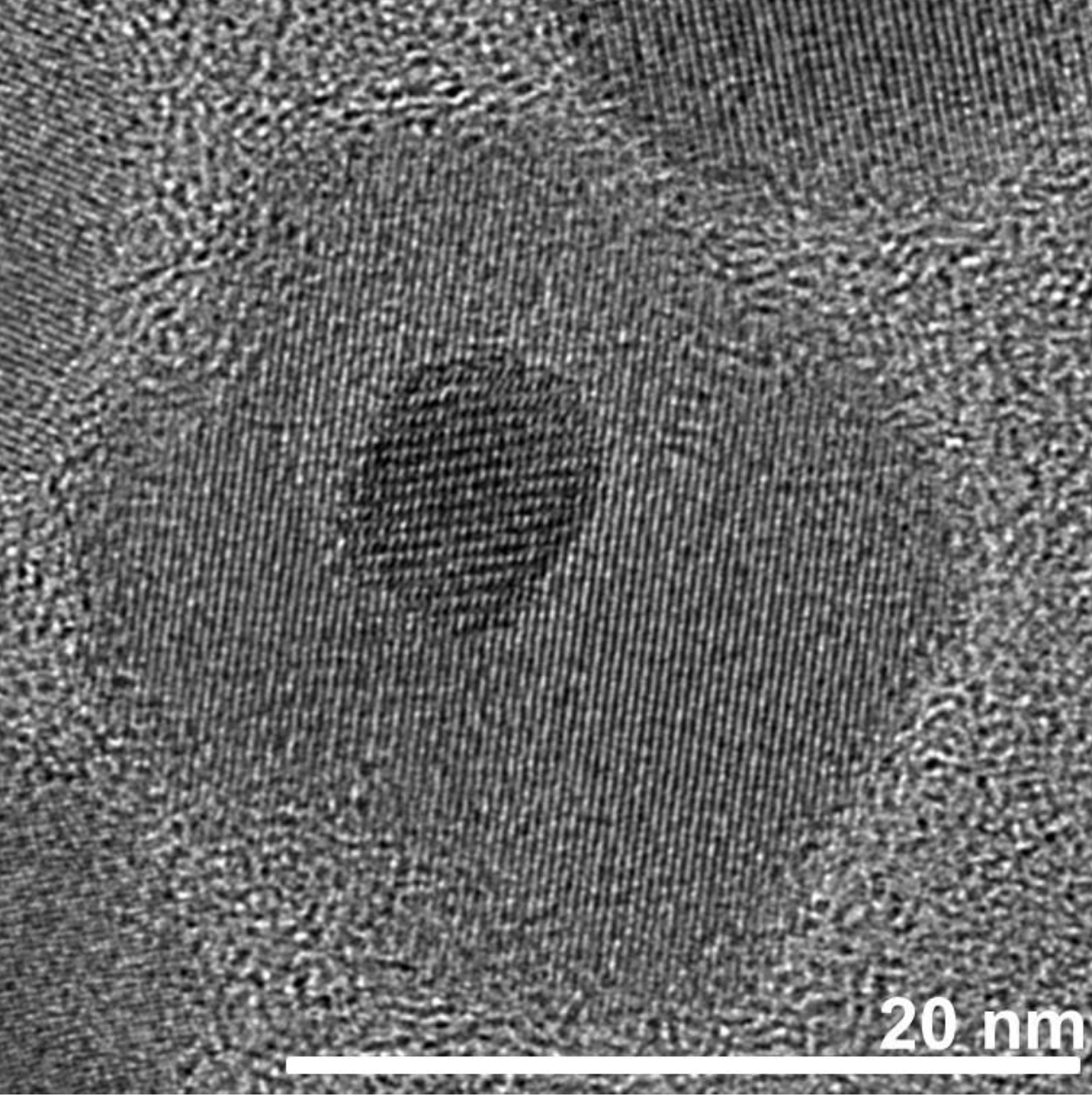}

\end{subfigure}%
\begin{subfigure}{.5\columnwidth}
  \centering
  \includegraphics[width=1.\columnwidth]{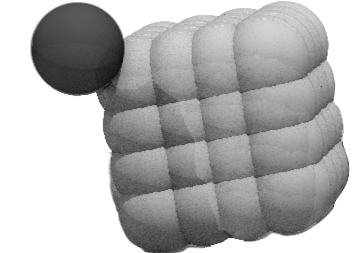}

\end{subfigure}
\caption{Left: HR-TEM images of the self-propelled dipolar cubes. In the bright-field images, darker contrast corresponds to platinum, while lighter contrast corresponds to cobalt ferrite. Right: computational "raspberry" model of the self-propelled dipolar cube. The dark-gray particle attached to the corner of the cube represents the active platinum sphere in the experimental system.}
\label{fig:exp}
\end{figure}

The investigated active particle consists of a single-domain cobalt ferrite ({CoFe}) nanocube with a smaller platinum (Pt) nanoparticle rigidly attached to one of its corners, as shown in Fig.~\ref{fig:exp}. The synthetic preparation of cobalt ferrite-platinum nanostructures entails a stable interface linkage between the domains via a two-step process \cite{zhang2008morphological}. Initially, platinum nanoparticles in a size range of $(6.4 \pm 0.8)$nm are prepared by a modified thermal decomposition route \cite{inprep}. The seed-mediated growth of the {CoFe} domain is induced by thermolysis and performed in organic phase nucleated from the platinum counterpart. Next, the capping surface of these nanostructures is further modified using a ligand exchange treatment. In this way, it is possible to obtain poly (acrylic acid) capped nanostructures with an edge length of $(29.6 \pm 3.7)$nm. These nanostructures are well-dispersible and stable in aqueous media as well as in diluted buffer solutions at physiological pH. The units are activated through self-diffusiophoresis using a coupled chemical fuel system based on the platinum catalysed reduction of borohydrides. Hence, the propulsion is generated by a concentration gradient considering both fuel and reaction products in the vicinity of the platinum counterpart. In addition, the dipolar CoFe units are railed by a passive magnetic field to reduce the predominant rotational diffusion affecting for nanometric objects.

The actual orientation of the magnetic moment within the cube and with respect to the corner where the active particle is attached is difficult to determine experimentally. The crystalline anisotropy suggests that [111] orientation is the most probable one. However, the orientation in real systems might differ or even switch in the course of the experiment. One of the ways to elucidate the impact of intrinsic magnetisation direction on the self-propulsion behaviour in an applied magnetic field is to employ simulations, in which, as shown below, it is straightforward to manipulate this parameter and analyse how it affects the efficiency of the external field on directing the active diffusion.

\subsection{Numerical model}\label{subs:sim}

We perform molecular dynamics simulations in the canonical ensemble to investigate the effects of the anisotropies in the active diffusion of the dipolar cube under infinite dilution conditions. All simulations were executed with the simulation package {ESPResSo} \cite{weik2019espresso}. The fundamental equation which is numerically integrated to get a discrete trajectory of the particle is the Langevin equation of motion, that for translational degrees of freedom reads
\begin{equation}\label{Langevin}
m\frac{d\textbf{v}}{dt}=-\gamma \textbf{v}-\nabla U(\textbf{r})+F(t) \, ; \, \textbf{v}=\frac{d\textbf{r}}{dt},
\end{equation}
including the quantities: particle mass $m$, particle position $\textbf{r}$, particle velocity vector $\textbf{v}$, friction coefficient $\gamma$, the gradient of any interaction potential acting on the particle $\nabla U(\textbf{r})$ and a random force $F(t)$. The latter should be Gaussian distributed according to Ornstein and Uhlenbeck \cite{uhlenbeck1930theory}, having independent components with magnitude $D_p$, and $\delta$-correlated time dependence,
\begin{equation}
\begin{split}
\langle F(t) \rangle &= 0, \, \\
\langle F_i (t) F_j (t^\prime) \rangle &= 2D_p \delta_{i,j}\delta(t-t^\prime)\,, 
\label{noise}
\end{split}
\end{equation}
where $i$ and $j$ can take the indices $x,y$ and $z$ of the spatial directions. This random force represents implicitly the effects of the thermal fluctuations of the background fluid at a temperature $T$, providing the possibility to efficiently produce simulation data for a system at constant thermal energy $kT$, being $k$ the Boltzmann constant. Expressions analogous to (\ref{Langevin}) and (\ref{noise}) are also integrated for the rotational degress of freedom.

The interaction of the cube magnetic moment, represented as a fixed point dipole $\vec d$, with the applied magnetic field, $\vec H$, is given by the classical Zeeman potential
\begin{equation}
U_Z=-\vec{d} \cdot \vec{H},
\end{equation}
In all cases, the field is applied along the $z$-axis of the simulation box, that is cubic and has periodic boundary conditions in order to represent a pseudo-infinite system.

Exploiting the Langevin-equation, self-propelled motion in simulations can be represented by a constant force applied along the propulsion axis of the active particle. Through a velocity dependent friction, an active particle attains a terminal velocity $v_a$ (see Table \ref{tab:parameters}). The value of the latter is defined by the balance of this friction and the driving force. This approach is called the Active-Brownian-Particle (ABP) model and is used to study a variety of active matter phenomena in simulations \cite{romanczuk2012active,shaebani2019computational}.

\begin{figure}[h!]
\vspace{-10pt}
\centering
\includegraphics[width=0.8\columnwidth,height=0.8\columnwidth]{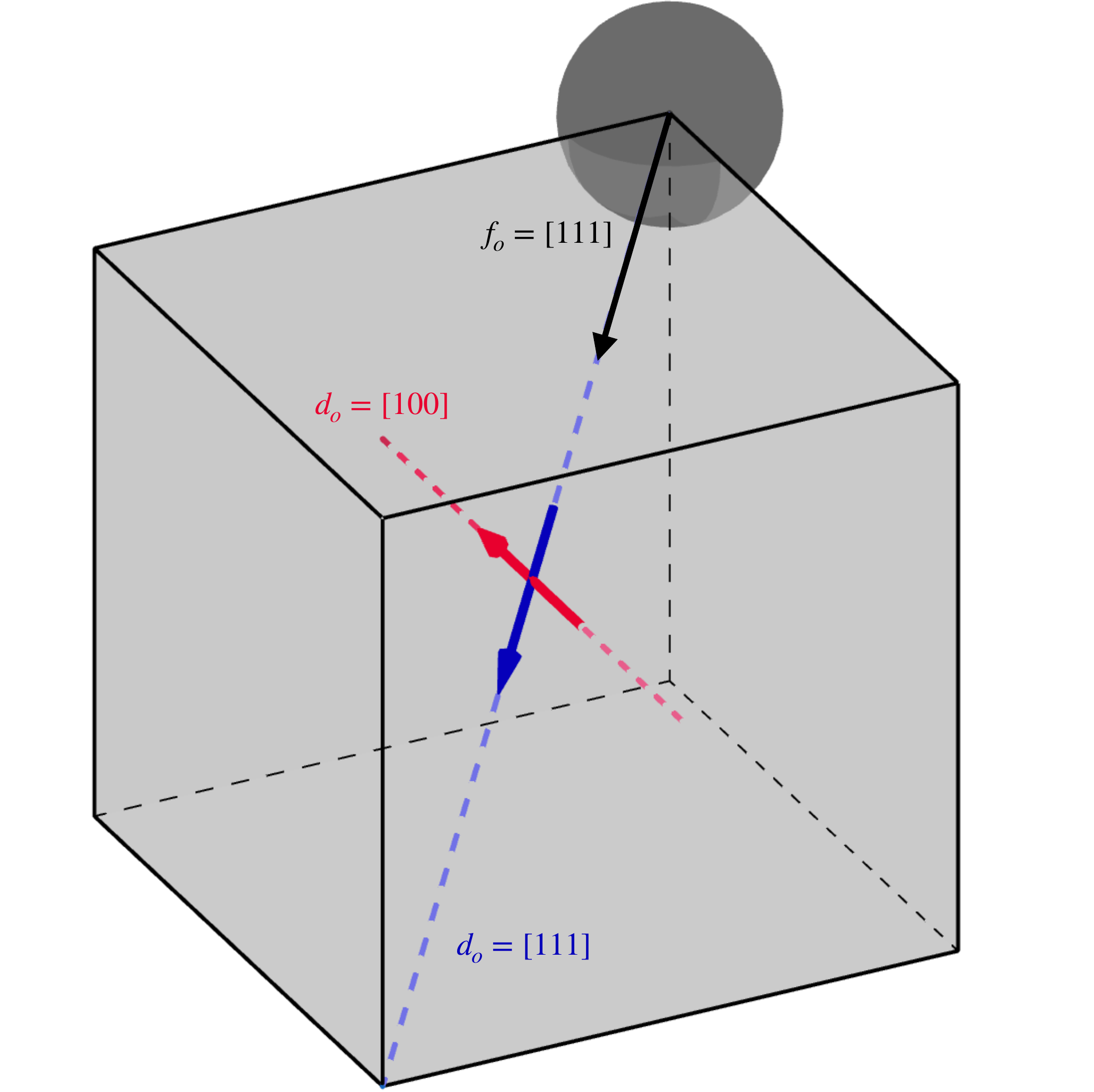}
\caption{Schematic of an active-cube unit, with the cube in light-gray and the active portion as a dark-gray sphere. The vectors in the center represent the point dipole in the center in the cube, which has a direction $d_o$ aligned with either the [111] (in blue) or the [100] (in red) axis of the cubes body. The black vector indicates the orientation $f_o=[111]$ of the induced active velocity, which is pointing towards the cube center at all times.}
\label{fig:schematic}
\end{figure}

\begin{table*}[h!]\centering
\ra{1.3}
\begin{tabular}{@{}cccccccccccc@{}}\toprule\midrule
& $\left | \vec{d}\, \right |$ & $\left | \vec H_{max} \right |$ &
 $k T$ & $\gamma$ & $m_{\mathrm{CoFe}}$ & $m_{\mathrm{Pt}}$ & $v_a$\\ 
 \midrule

& 10 & 0.1 & 1 & 1 & 56 & 2 & 0.1\\
\bottomrule
\end{tabular}
\caption{Dimensionaless parameters used in the simulations: dipole moment of the magnetic cube, $\left | \vec{d}\, \right |$; maximum strength of the applied field, $\left | \vec H_{max} \right |$; thermal energy, $k T$; friction coefficient, $\gamma$; mass of the magnetic cube, $m_{\mathrm{CoFe}}$; mass of the catalytic sphere, $m_{\mathrm{Pt}}$; terminal velocity of the active force, $v_a$. See the main text for their correspondence to experimental values.}
\label{tab:parameters}
\end{table*}

The approach used to represent the anisometry of our system is the so-called ``raspberry model'' \cite{de2015raspberry,fischer2015raspberry,lobaskin2004new}, which uses spherical building blocks disposed in a rigid arrangement to effectively represent any body shape. This is analogous to the model used by Donaldson \textit{et al.} for simple magnetic nanocubes \cite{donaldson15a,2017-donaldson-acsn,rossi2018self}. The size, shape and mass of the raspberry grains are chosen in such a way, that the resulting cube characteristics closely resemble those in experiment. The number of grains in the raspberry is not directly affecting the diffusion of the particles, but only its geometry. It is important to mention that once hydrodynamics is taken into account, the choice of size and number of raspberry grains becomes undeniably important. Fig.~\ref{fig:exp} shows the raspberry structure used in our simulations. The point dipole is placed in the center of the cube. Its relative orientation $d_o$, defined with respect to the corner at which the active particle is fixed, is either [111] (along one main diagonal of the cube, pointing opposite to the reference corner) or [100] (pointing to the center of one of the faces of the cube that includes the reference corner). This is depicted in the schematic representation shown in Fig.~\ref{fig:schematic}. The self-propulsion force of the active sphere sitting at the reference corner is constant and pointing towards the center of the cube, thus, parallel to the dipole moment for the case [111] and perpendicular to it for [100]. The analysis of the simulation trajectories is performed separatelly for the direction parallel ($z$-axis) and perpendicular ($xy$-plane) to the applied field.

In all simulations the integration time step has been chosen to be $\tau=0.005$, ensuring stability of the integration scheme. The simulation protocol is the following. First, a cycle of $2000\tau$ to allow the random reorientation of the raspberry particle from its initial configuration has been performed before the magnetic field is switched on along the $z$-axis. Another cycle of the same length provides time for the unit to respond to the magnetic field. The mean-squared displacement (MSD) is calculated after such initial relaxations, for at least $10^7\tau$ steps, to provide sufficient statistics for the range of the MSD investigated in this study. Results below were obtained by averaging over 4 independent simulation runs.

\subsection{Connection to the experiments}\label{subs:params}
The parameters used in the model represent CoFe cubes of 20~nm side length with a magnetic moment of $1.05 \cdot 10^{-18}\,\mathrm{A\cdot m}^2$ attached to Pt spheres of 4~nm of diameter. The highest strength of the magnetic field is chosen so that the absolute value of the minimum Zeeman energy, corresponding to a perfect aligment of the dipole with the field, is equal to the thermal energy. This corresponds to a field of approximately 3.9~mT. However, in order to ensure the stability of the numerical integrations, in simulations it is convenient to use a system of dimensionless units that gives parameter values around unity. That system can be arbitrary as long as it keeps the same ratios for the corresponding relevant experimental parameters. In our case, we set our dimensionless parameters by considering the experimental ratios that have been already determined for this system \cite{inprep}. These are the aforementioned ratio between magnetic and thermal energy, $|U_Z|/k_bT=1$, and the mass ratio of the Pt and CoFe components, $m_{\mathrm{CoFe}}/m_{\mathrm{Pt}}=28$. Since here we focus only on qualitative results, arbitrary values have been chosen for parameters not yet measured experimentally. Table~\ref{tab:parameters} summarises the dimensionaless values used in this study. The correspondence between the experimental and computational time is not useful due to the absence of explicit hydrodynamics. However, this does not qualitatively affect the main results reported below.

\section{Results and Discussions}\label{ref:rnd}

\subsection{Mean-squared displacement and diffusion}\label{subs:msd}
\begin{figure}[h!]
\centering
\includegraphics[width=0.8\columnwidth]{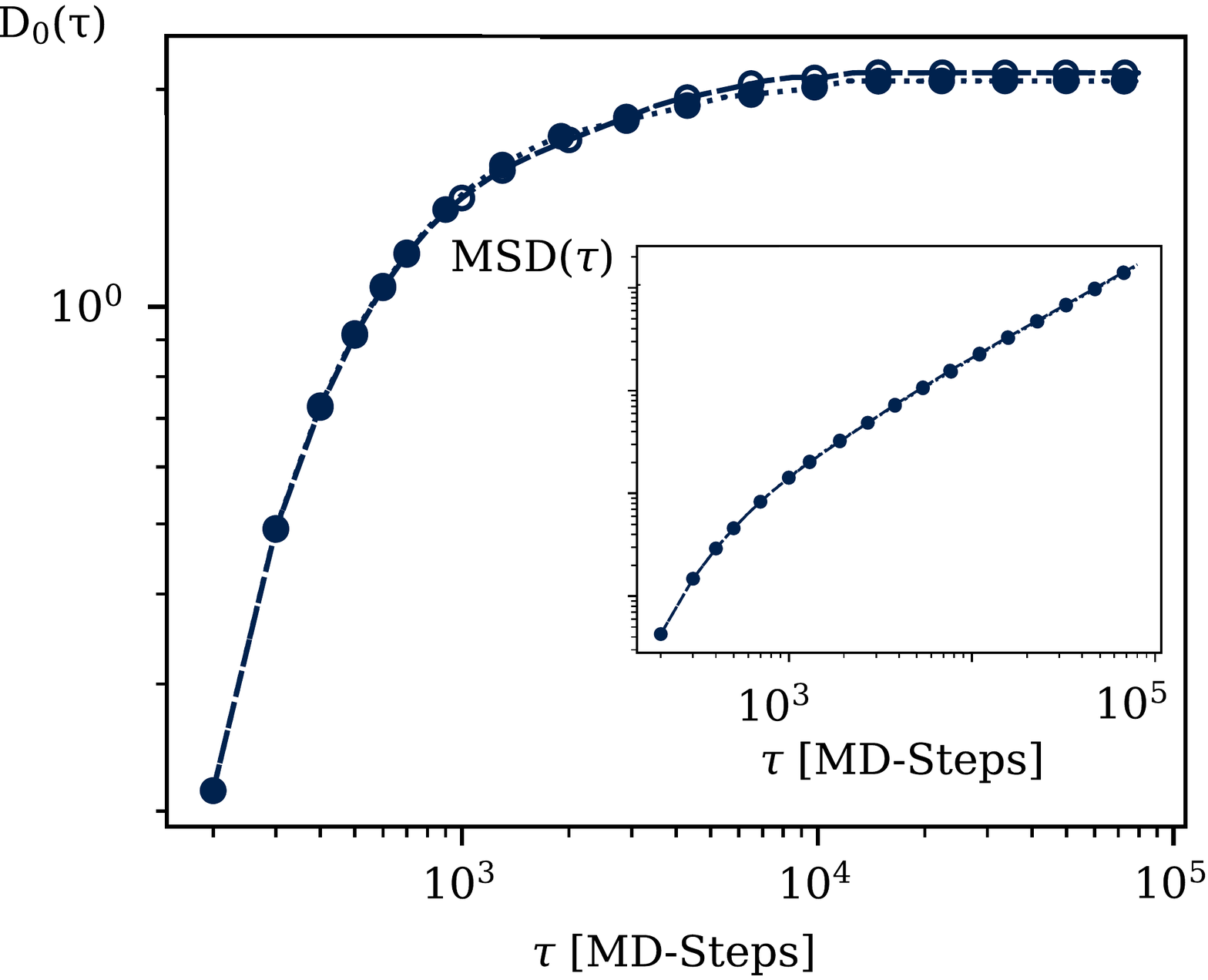}
\caption{Diffusion of an unconstrainted active particle without the cube attached, obtained from the computation of the MSD divided by $\tau$. Dashed lines and empty markers correspond to the component parallel to the magnetic field, $D_0^\parallel$, dotted lines and filled markers to the mean of the two perpendicular components, $D_0^\perp$.}
\label{fig:D0}
\end{figure}

In order to quantify diffusion anisotropy, we investigated the MSD and diffusion coefficient $D_\tau$ for a range of applied field strengths up to $|U_Z|/k_bT=1$ for a single active unit to see how much its diffusion is influenced by the magnetic field. As a reference, the diffusion $D_0(\tau)$ of only the active particle without the cube, but otherwise same parameters, is computed by dividing the MSD by $2 \tau$ per component. The calculation of the diffusion is split into two parts, with the component $D_0^\parallel$ being the diffusion along the field direction and the component $D_0^\perp$ being the mean of the diffusion components perpendicular to the field. The diffusion, shown in Fig.~\ref{fig:D0}, follows a typical behaviour of a free, unconstrained active particle as in Ref.~\cite{bechinger2016active}, reaching constant saturation value after a ballistic regime. In this short time ballistic regime, the particle can move in a more or less straight line before it is reoriented by random kicks from its surrounding medium. The duration of this regime is determined by the strength of the active force acting on the particle. The fact that both curves are almost identical evidences a fully isotropic diffusion, as expected.

\begin{figure}[t!]
\centering
\includegraphics[width=0.85\columnwidth]{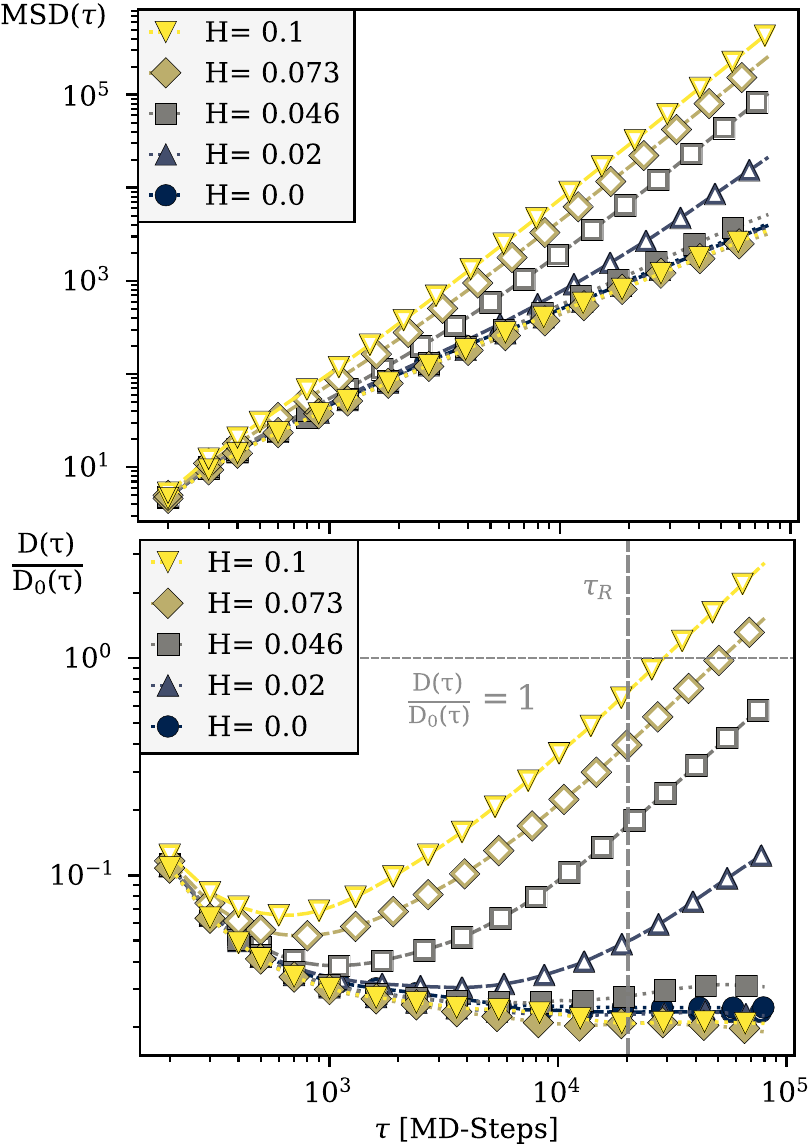}
\caption{MSD (upper) and $D(\tau)$ (lower) of a single active cube unit in fields with magnitude as marked by colours and symbol shapes, where the dipole orientation $d_o$ lies along the [111] axis of the cube. Dashed lines and empty symbols correspond to the component parallel to the magnetic field, dotted lines and filled symbols to the mean of the two perpendicular components. The vertical bold line indicates $\tau_R$ and the horizontal thin line the point where $D(\tau)/D_0(\tau)$ is equal to 1.}
\label{fig:111}
\end{figure}

Top panel in Fig.~\ref{fig:111} shows the split MSD of the whole hybrid active-magnetic particle with orientation of the magnetic moment [111] for different field strengths, whereas the bottom panel shows its split diffusion coefficients divided by the corresponding reference value of the free active sphere, $D^\parallel/D_0^\parallel$ and $D^\perp/D_0^\perp$. Here, dotted lines and filled symbols are the average of the two components perpendicular to the field and dashed lines with empty symbols are the parallel components. To guide the eye, the point $\tau_R$ after which all perpendicular components reached constant diffusion, $(dD(\tau)/d\tau)_{\tau \ge \tau_R}=0$, and the ratio of perturbed and non perturbed diffusion equal to unity, $D(\tau)/D^0(\tau)=1$, are indicated by dashed horizontal and vertical lines, respectively. The MSD reveals that the effect of particle redirection stemming from the interaction of the dipolar cube with the magnetic field leads to a significant enhancement of the parallel component of the displacement, with an increasing slope at stronger fields. This is equally reflected in the diffusion coefficient, where it becomes clear that the perpendicular to the field components saturate to a state of constant diffusion while the parallel components do not follow this behaviour and keep growing beyond the investigated time frame. Additionally, the perpendicular components saturate at a more than 10 times smaller diffusion coefficient than its unperturbed counterpart, $D_0$. Besides not saturating, the parallel components are also only smaller than their counterpart for short time scales, eventually increasing over the point where $D(\tau)/D_0(\tau)=1$. The time at which $D(\tau)$ and $D_0(\tau)$ are equal, depends on the strength of the applied field, and is shorter as the magnetic field strength increases.

\begin{figure}[t!]
\centering
\includegraphics[width=0.85\columnwidth]{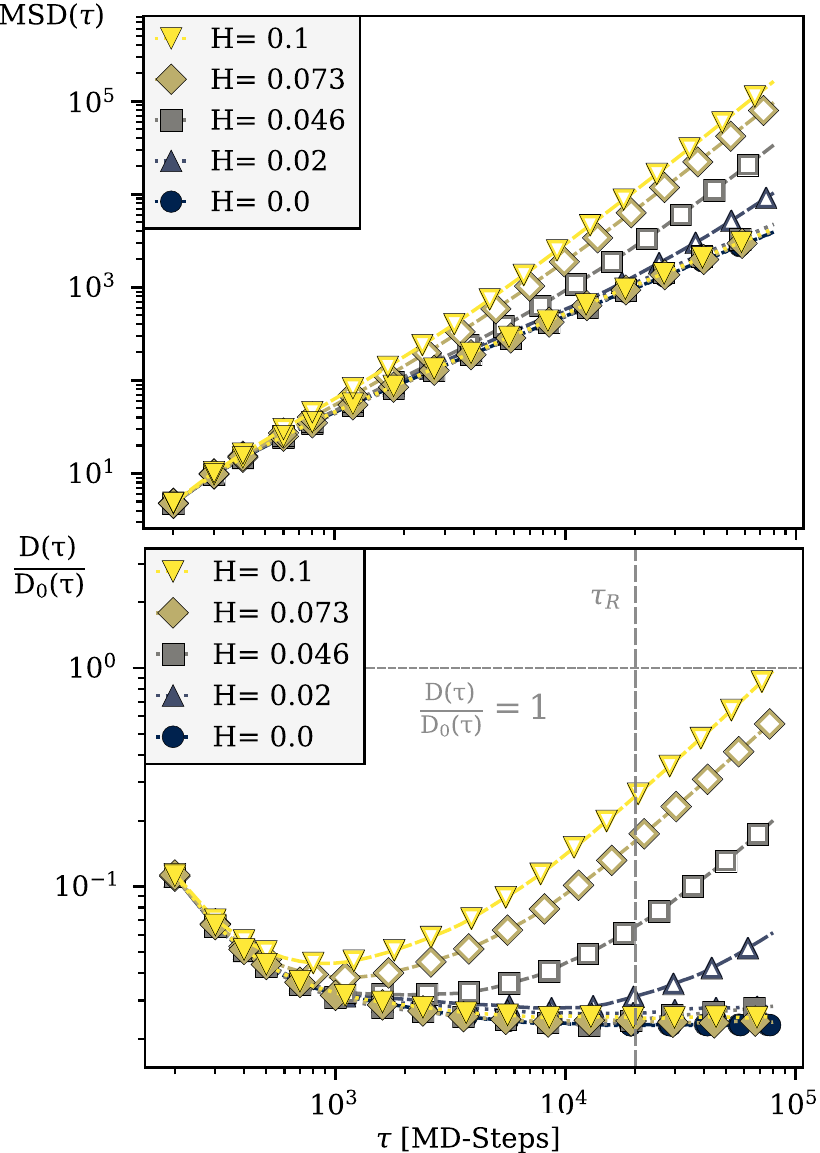}
\caption{MSD (upper) and $D(\tau)$ (lower) of a single active cube unit in fields with magnitude as marked by colours and symbol shapes, where the dipole orientation $d_o$ lies along the [100] axis of the cube. Dashed lines and empty symbols correspond to the component parallel to the magnetic field, dotted lines and filled symbols to the mean of the two perpendicular components. The vertical bold line indicates $\tau_R$ and the horizontal thin line the point where $D(\tau)/D_0(\tau)$ is equal to 1.}
\label{fig:100}
\end{figure}

Fig.~\ref{fig:100} shows the same quantities as the just discussed Fig.~\ref{fig:111}, but for the case of dipole orientation $d_o=[100]$. Qualitatively, the behaviour stays intact, but quantitative changes are observed. Within the same time-frame, the diffusion does not grow as high as in the [111] case, and only the parallel component for a magnetic field where $|U_Z|/k_bT=1$, manages to reach the point where $D(\tau)$ and $D_0(\tau)$ are equal. This is attributed to the unfavorable configuration of the active force compared to the dipole orientation. The dipole remains mostly aligned with the magnetic field -- as shown in later parts of this section -- and the active force can not anymore propagate its full energy onto the effective diffusion direction of the cube unit, resulting in an overall slower increase of the diffusion over time as compared to the more favourable [111] case.

It may seem surprising that the ratio of in field components keep growing with increasing $\tau$ in Figs.~\ref{fig:111}  and~\ref{fig:100}. The reason for this growth is that the distance, traveled by the active cube in the direction of the applied field, is growing infinitely large with increasing time step, in contrast to Fig.~\ref{fig:D0}, where thermal fluctuations force the unbiased active particle to perform random motion, leading to the saturation of $D_0(\tau)$. One can understand this by imagining a particle that is moving in one direction with a constant velocity. If you measure the distance the particle traveled during certain time frames, one will see that the distance travelled during larger times is bigger than for smaller ones.

\begin{figure}[h!]
\centering
\includegraphics[width=0.85\columnwidth]{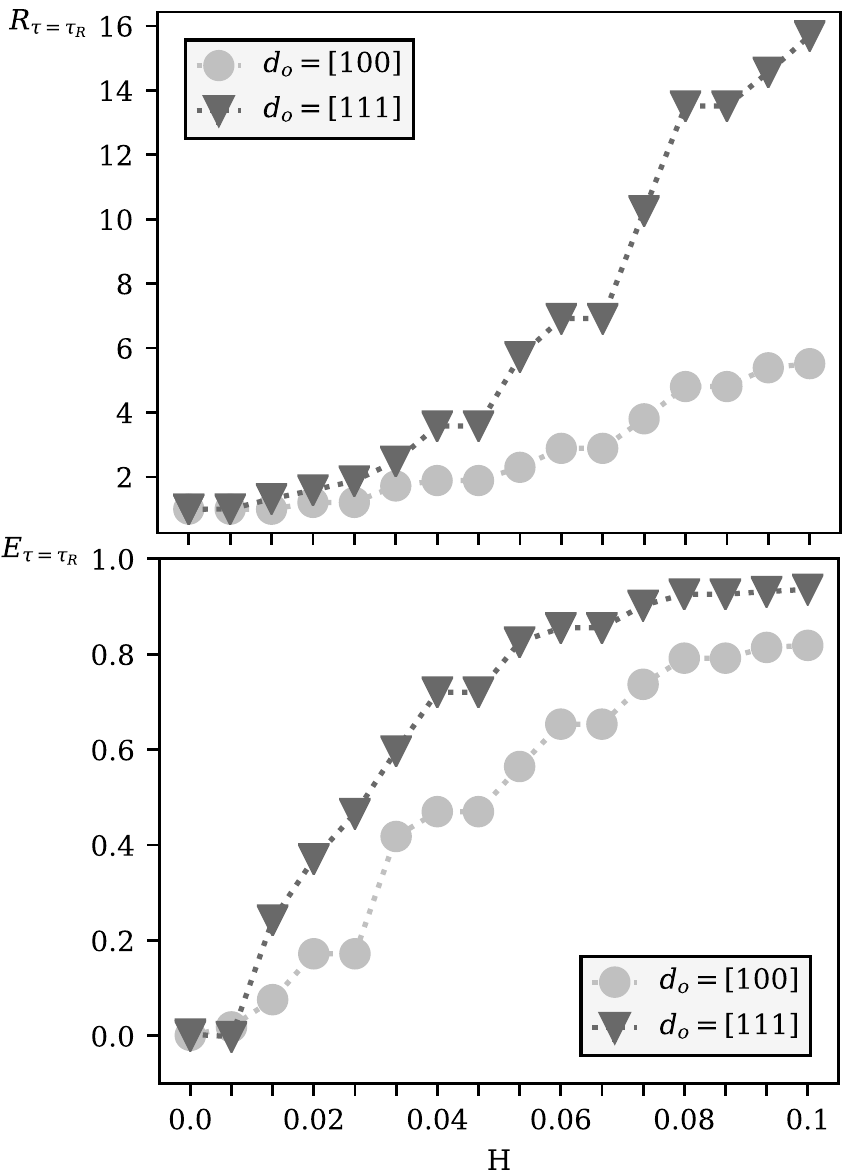}
\caption{Ratio (upper) and transport-efficiency (lower) versus H, computed for the values of $D_\parallel$ and $D_\perp$ at $\tau=\tau_r$. The results for the dipole orientation $d_o=[100]$ are shown with circles and for $d_o=[111]$ with triangles. Dotted lines are guides to the eye.}
\label{fig:eff}
\end{figure}

At this point we define two additional quantities useful for the discussion: the ratio $R=(D_\parallel/D_\perp)_{\tau=\tau_R}$ of the parallel and perpendicular components of the diffusion at point $\tau=\tau_R$, and the transport efficiency $E=1-(D^\perp/D^\parallel)_{\tau=\tau_R}$. The dependence of these parameters on the field strength for both dipole orientation is shown in the upper and lower panels of Fig.~\ref{fig:eff}, respectively. Here one can see that, within the investigated range of magnetic field strengths, for the [111] case, diffusion in direction parallel to the field is up to 16 times higher than diffusion of the perpendicular components and the efficiency therefore goes up to ~94\%. For the [100] case, although not all energy of the active particle is distributed towards diffusion in field direction, the parallel component is still up to 5 times bigger in the same field strength range, topping out at ~80\% efficiency. The above measures disclose the ability of the magnetic field to direct the active cube and enhance its diffusion along its direction. So far unclear, taking into account the geometry of the cube, is whether diffusion happens parallel or anti-parallel to the field direction, as the diffusion coefficients alone are not suitable to make a precise statement about this point.

\subsection{Angular anisotropy and trajectories}\label{subs:ang}

\begin{figure*}[h!]
\centering
\includegraphics[width=0.9\textwidth]{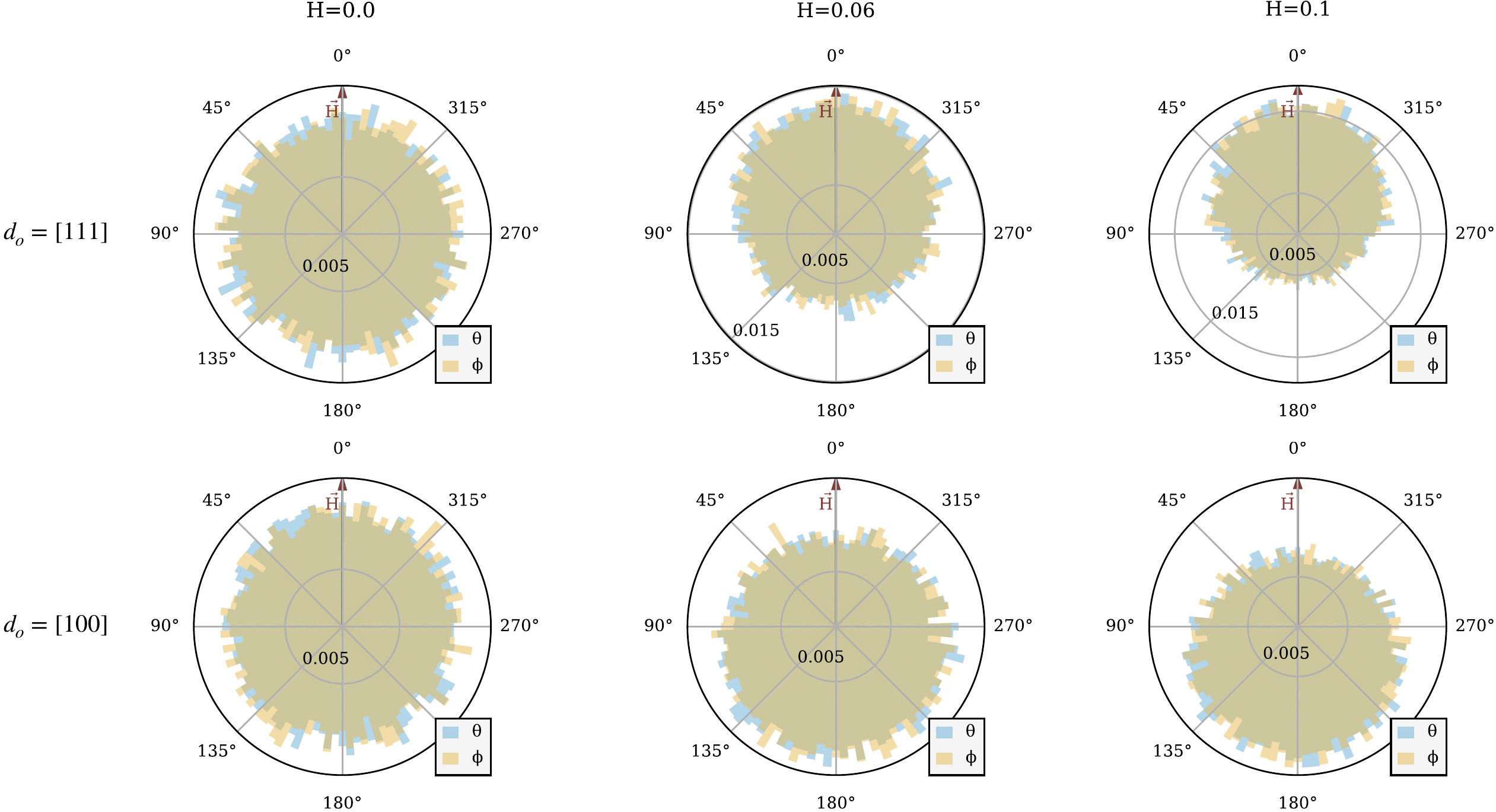}
\caption{Polar distributions of the track-angle -- the angle enclosed between a vector from one trajectory point to another and the magnetic field vector ($\Vec{H}$, in red, aligned with the \ang{0} line) --  for the zx-plane component ($\theta$, in light-blue) or zy-plane component ($\phi$, in light-yellow). The row of the figures determines the dipole orientation withing the cube, whichs is either [111] in the upper row or [100] in the lower row. The columns determine the strength of the applied field ranging from H=0.0 on the left, to H=0.1 on the right.}
\label{fig:pol}
\end{figure*}

In order to elucidate the question rised above, in Fig.~\ref{fig:pol}, we present  polar relative probability distributions of the track-angle -- \textit{i.e.}, the angle between the vector connecting two consequent points of the trajectory and the magnetic field vector. To have a deeper insight into possibly occurring anisotropies, the analysis of this distribution is split into the angle that the projection of the vector trajectory onto the $xy$-plane encloses with the magnetic field, and the angle that the corresponding projection onto the $yz$-plane encloses with the latter. Here, those two angles are denoted with $\phi$ and $\theta$, respectively. Probability distributions for the values of these angles obtained for three selected field strengths, $H=0.0,0.06,0.1$, are displayed in these plots. The distributions are a representation of the mean direction traveled by the particle with respect to the magnetic field. It is seen that fields of increasing magnitude clearly shift the distribution towards the \ang{0} axis for the [111] case, showing a clear deviation from the uniform distribution corresponding to an unperturbed active cubic particle at 0 field. Both, $\phi$ and $\theta$, follow this pattern. The [100] case shows an strikingly opposite behaviour. Here, the particle travels in the opposite direction of the magnetic field. This observation can be explained in two steps. First, the angle between the active force and the [100] dipole is acute enough, so that the active particle is effectively pushing in an opposed direction to which the dipole is pointing (refer back to Fig. \ref{fig:schematic} as a visual aid). Second, the force of the active particle is not strong enough to fully break the alignment of the dipole and the magnetic field. Those two factors force the active particle to push the unit towards the counter-direction of the dipole, hence the direction of the magnetic field, and we therefore get the distributions observed in Fig.~\ref{fig:pol}. Furthermore, the distributions of the [111] case are more narrow than the ones for the [100] case at the same field strength. This stems from the same reason as the magnitude difference in the diffusion coefficient graphs for the two cases: the active particle is not able to fully project its energy onto the effective diffusion direction. The track-angle distribution is a quantity accessible in experimental set-ups and can therefore act as a guide to determine the internal orientation of the synthesised CoFe cubes.
\section{Conclusion}\label{sec:con}
By performing Langevin Dynamics simulations, we investigated the behaviour of a hybrid magnetic nanocube with a smaller active catalytic particle attached to one of its corners, when placed under the influence of an applied homogeneous magnetic field. We found that fields at which the Zeeman energy is equal to the strength of the thermal fluctuations are sufficient to drastically orient the cube along the magnetic field. This stems from the inability of the active component to overpower the reorientation process due to the magnetic field, even if the active force is not directly aligned with the direction of the dipole. The active diffusion parallel to the field is growing slower for the case of non-aligned active force-dipole orientations, as the active force is not fully propagated along the diffusion direction of the particle and some energy is therefore lost. The direction of diffusion is determined by the internal magnetisation orientation, showcased for two orientations along the [111] and [100] axis of the cube. The [111] case diffuses parallel to the magnetic field, while the [100] particle is geometrically constrained to move anti-parallel to the field. This can act as a guide to ascertain the magnetisation orientation of synthetically created magnetic cubes, as the observation of this behaviour can also be captured experimentally, and may furthermore act as a sorting procedure to separate cubes with different magnetisation orientations if an active force on the units is present. Regardless, both scenarios are highly suitable for potential applications where field induced redirection can be beneficial, as the transport parallel to the field is up to 16 times higher for the [111] and up to 5 times higher for the [100] case. Being these results only for magnetic fields where the Zeeman energy and thermal fluctuations are of the same order, we expect a much higher efficiency of the process at stronger fields, which can still be easily achieved in experimental set-ups.

\section*{Acknowledgements}
This research has been supported by the Russian Science Foundation Grant No.19-12-00209. Authors also acknowledge support from the Austrian Research Fund (FWF), START-Projekt Y 627-N27. Y.M. gratefully acknowledges a Doctoral Scholarship granted by the Deutscher Akademischer Austauschdienst (DAAD). A.M.S. acknowledges funding from DFG-SPP 1681, grant number SCHM1747/10. Computer simulations were performed at the Vienna Scientific Cluster (VSC-3).
\bibliographystyle{elsarticle-num-names}
\bibliography{biblio.bib}

\end{document}